\documentclass[twocolumn,pra,showpacs,amsmath,amssymb,superscriptaddress,
			   nofootinbib
			   ]{revtex4-2}
\usepackage{graphicx}
\usepackage{amssymb}
\usepackage{tikz}
\usetikzlibrary{shapes.arrows, shapes.geometric, arrows, spy}
\usepackage{diagbox}
\usepackage{layout}
\usepackage{mathtools}
\usepackage{url}
\usepackage[hidelinks, colorlinks=true, linkcolor=blue, citecolor=blue, urlcolor=blue]{hyperref}
\usepackage[caption=false]{subfig}
\tikzstyle{startstop} = [rectangle, rounded corners, minimum width=3cm, minimum height=1cm, text centered, draw=black, fill=red!30]
\tikzstyle{numerical} = [rectangle, minimum width=3cm, minimum height=1cm, text centered, draw=black, fill=blue!30] 
\tikzstyle{analytical} = [rectangle, minimum width=3cm, minimum height=1cm, text centered, draw=black, fill=green!30] 
\tikzstyle{decision} = [diamond, minimum width=3cm, minimum height=1cm, text centered, draw=black, fill=yellow!30]
\tikzstyle{arrow} = [thick,->,>=stealth]

\usepackage{comment}

\begin{document}
	
	\title{
		Local models and Bell inequalities for the minimal triangle network
		}
	\author{Jos\'e M\'ario da Silva}
	\email{jose.filho@ufpe.br}
	\affiliation{Departamento de
		F\'{\i}sica, Centro de Ci\^encias Exatas e da Natureza, Universidade Federal de Pernambuco, Recife, Pernambuco
		50670-901 Brazil}
	\author{Alejandro Pozas-Kerstjens}
	\email{physics@alexpozas.com}
	\affiliation{Department of Applied Physics, University of Geneva, 1211 Geneva 4, Switzerland}
	\author{Fernando Parisio}
	\email{fernando.parisio@ufpe.br}
	\affiliation{Departamento de
		F\'{\i}sica, Centro de Ci\^encias Exatas e da Natureza, Universidade Federal de Pernambuco, Recife, Pernambuco
		50670-901 Brazil}
	
	\begin{abstract}
		Nonlocal correlations created in networks with multiple independent sources enable surprising phenomena in quantum information and quantum foundations. The presence of independent sources, however, makes the analysis of network nonlocality challenging, and even in the simplest nontrivial scenarios a complete characterization is lacking. In this work we study one of the simplest of these scenarios, namely that of distributions invariant under permutations of parties in the minimal triangle network, which features no inputs and binary outcomes. We perform an exhaustive search for triangle-local models, and from it we infer analytic expressions for the boundaries of the set of distributions that admit such models, which we conjecture to be all the tight Bell inequalities for the scenario. Armed with them and with improved outer approximations of the set, we provide insights on the existence of a classical-quantum gap in the triangle network with binary outcomes.
	\end{abstract}
	\maketitle
	
	The incompatibility of quantum mechanics with local hidden variable theories, predicted by Bell \cite{Bell} and confirmed by increasingly convincing experiments \cite{experimental1,experimental2,experimental2b,experimental3,experimental4,experimental5}, represents a fundamental departure from classical intuition. Beyond its foundational implications, Bell nonlocality \cite{revision_bell} is now recognized as a key resource for quantum technologies \cite{Ekert91,Pironio2010,DI,selftesting}.
	
	A natural generalization of Bell's scenario are networks \cite{revision_networks} where several independent sources distribute systems to multiple parties.
	These architectures naturally appear, e.g., in quantum communication \cite{satellite1} and in schemes for quantum state transfer \cite{CommQuantNet1,CommQuantNet2}.
	Thus, the characterization of classical and quantum correlations in networks is fundamental to identify quantum advantage in these and other applications. However, the independence of the sources makes these sets of correlations to be non-convex, and thus difficult to characterize.
	Despite this, many methods have been developed to achieve partial characterizations and develop non-locality witnesses \cite{correlation_scenarios,bilocality_extended,inflation,entropy,cov}.
	
\begin{figure}
		\centering
		\subfloat[\label{fig:triangle scenario}]{
				\scalebox{0.8}{\begin{tikzpicture}[scale=0.85,
					output/.style={circle, draw=black!60, fill=black!5, very thick, minimum size=10mm},
					hidden/.style={inner sep=0,minimum size=5mm},
					input/.style={rectangle,draw=blue!60, fill=blue!5, very thick, minimum size=6mm}
					]
					\node[output] (a) at (0,0) {$A$};
					\node[hidden] (beta) at (60:2) {$\beta$};
					\node[hidden] (gamma) at (2,0) {$\gamma$};
					\node[output] (b) at (4,0) {$B$};
					\node[output] (c) at (60:4) {$C$};
					\path (60:4)+(-60:2) node[hidden] (alpha) {$\alpha$};
					
					\draw[-stealth] (beta) -- (a);
					\draw[-stealth] (beta) -- (c);
					\draw[-stealth] (gamma) -- (a);
					\draw[-stealth] (gamma) -- (b);
					\draw[-stealth] (alpha) -- (b);
					\draw[-stealth] (alpha) -- (c);
					
					\node[single arrow, draw, rotate=270, minimum height=0.6cm, minimum width=0.4cm, single arrow head extend=0.08cm] at (0, -1) {};
					\node at (0, -1.8) {$\{-1,1\}$};
					\node[single arrow, draw, rotate=270, minimum height=0.6cm, minimum width=0.4cm, single arrow head extend=0.08cm] at ([shift=(60:4)] 0, -1) {};
					\node at ([shift=(60:4)]0, -1.8) {$\{-1,1\}$}; 
					\node[single arrow, draw, rotate=270, minimum height=0.6cm, minimum width=0.4cm, single arrow head extend=0.08cm] at ([xshift=4cm] 0, -1) {};
					\node at ([xshift=4cm] 0, -1.8) {$\{-1,1\}$};
				\end{tikzpicture}
			}
		}
		\hfill
		\subfloat[\label{fig:3d}]{
			\scalebox{0.99}{\includegraphics[width=0.5\columnwidth]{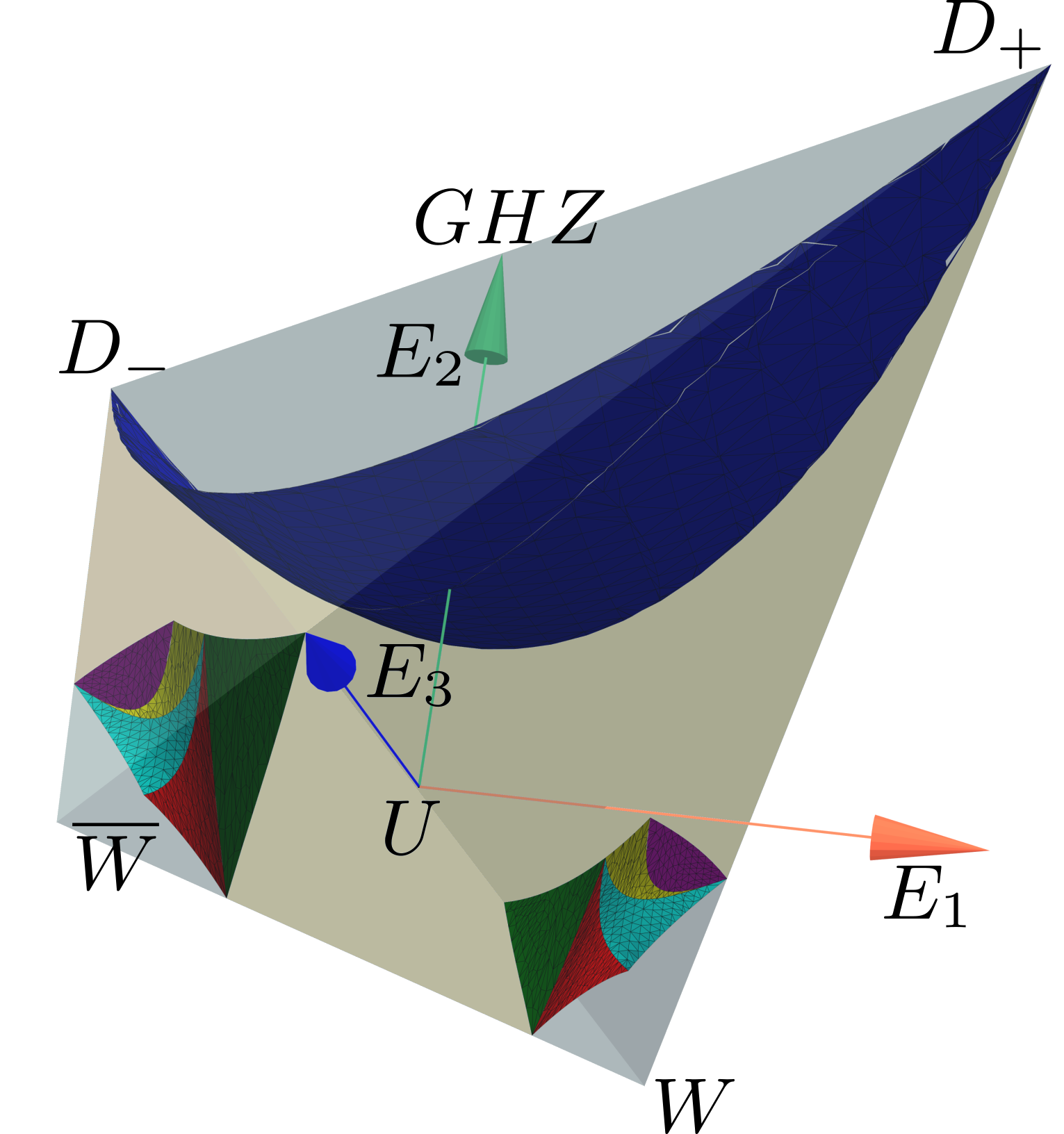}
			}
		}
		\caption{(a) The minimal triangle scenario. The parties $A$, $B$ and $C$ have no choice of inputs, and binary outputs $a, b, c\in\left\{-1, 1\right\}$.
		The sources $\alpha$, $\beta$, and $\gamma$ are independent. (b) A representation in correlator space of the symmetric, network-local distributions that can be generated in the minimal triangle scenario. One of our main contributions is an analytic characterization of the colored boundaries.
		The labeled distributions are defined in the main text.}
		\label{fig:intro}
	\end{figure}

	Among the possible networks, the triangle network [see Fig.~\hyperref[fig:triangle scenario]{1(a)}] has received considerable attention, since it is the simplest network where all parties are connected to each other.
	It supports genuine quantum nonlocality \cite{RGB4} and randomness certification without measurement choices \cite{Sekatski2023} (as opposed to the Bell scenario where this is impossible).
	However, the absence of marginal independences also makes it a challenging scenario to study, even in the simplest case without measurement choices and with binary-valued outcomes.
	There, the search for realizations is based on brute force \cite{numerical,neural_network}, while outer approximations are obtained via inflation methods \cite{inflation,quantum_inflation,NSI}.
	Using these tools, it has been discovered that the minimal triangle scenario supports supra-quantum correlations \cite{pozas2023minimal}, and many witnesses of non-locality have been given \cite{inflation_violations,pozas2022proofs}.
	Yet, a more relevant question is whether this scenario supports \textit{quantum} nonlocality, i.e., nonlocal distributions that can be generated by measuring quantum systems.
	Such correlations could enable, for instance, simple randomness generation protocols, while refuting them would redirect efforts to more promising scenarios.
	The question of the existence of quantum nonlocality in the triangle network with binary outcomes was posed in Ref.~\cite{inflation_violations}, and remains open to date.

In this work we provide an important step towards solving this question, by exhaustively analyzing the probability distributions symmetric under permutations of parties that can be created in the triangle network with no input choices and binary outcomes via local hidden variables. This analysis reveals that the upper bound in the cardinality of the hidden variables in Ref.~\cite{cardinality_bound} is not needed for a complete characterization of this set of distributions. Moreover, we find analytic expressions for the local boundaries. These are polynomial inequalities that, based on the strong numerical evidence built, we conjecture to be the tight Bell inequalities characterizing the scenario, depicted in Fig.~\hyperref[fig:3d]{1(b)}. Armed with them, and with outer approximations based on inflation \cite{inflation,quantum_inflation}, we provide insights on the potential existence of a classical-quantum gap in the binary-outcome triangle network: We shrink some previously known candidate regions for quantum nonlocality, while also identifying new ones where it could be found.

	\paragraph*{Local correlations in the minimal triangle.} We consider tripartite distributions $p(a,b,c)$ over binary outcomes $a,b,c\in\{-1,1\}$ that are invariant under party permutations.
	These distributions are completely described by the three symmetrized correlators $E_1 = \left\langle a \right\rangle = \left\langle b \right\rangle = \left\langle c \right\rangle$, $E_2 = \left\langle ab \right\rangle = \left\langle bc \right\rangle = \left\langle ac \right\rangle$, and $E_3 = \left\langle abc \right\rangle$, so that the probability distribution is
	\begin{equation}
		p(a,b,c)=\frac18 [1+E_1(a+b+c) + E_2(ab+bc+ca) + E_3abc].
		\label{eq:p}
	\end{equation}
	Due to the positivity constraints, valid behaviors lie in a tetrahedron in $(E_1,E_2,E_3)$ space [see Fig.~\hyperref[fig:3d]{1(b)}].

	A behavior is network-local if and only if there exist hidden variable distributions $q(\alpha)$, $r(\beta)$ and $s(\gamma)$, and response functions $A(a|\beta,\gamma)$, $B(b|\gamma,\alpha)$ and $C(c|\alpha,\beta)$, such that
	\begin{equation}
		\begin{aligned}
			p(a,b,c) = \sum_{\alpha=1}^{c_\alpha}\sum_{\beta=1}^{c_\beta}\sum_{\gamma=1}^{c_\gamma} &\,q(\alpha)r(\beta)s(\gamma)\\
			&\times A(a|\beta,\gamma)B(b|\gamma,\alpha)C(c|\alpha,\beta).
		\end{aligned}
		\label{eq:local}
	\end{equation}
	Here, $c_\alpha$, $c_\beta$ and $c_\gamma$ are the hidden variable cardinalities. Reference~\cite{cardinality_bound} showed that cardinalities $c_\alpha = c_\beta = c_\gamma = 6$ suffice to generate any local distribution in the minimal triangle network. Yet, it is not known whether such cardinalities are necessary.
	One of our main results is that, when restricting to distributions of the form of Eq.~\eqref{eq:p}, these maximum cardinalities can be reduced.

	\paragraph*{Local model search.} The search for a triangle-local model as in Eq.~\eqref{eq:local} can be written as a non-convex optimization problem, namely minimizing the distance between the target and modeled distributions.
	We implement this problem using the optimization routines in the \textsc{Python} package \texttt{scipy} \cite{scipy} (see the computational appendix \cite{computationalappendix}), and search for models with cardinalities up to the maximum $c_\alpha\,{=}\,c_\beta\,{=}\,c_\gamma\,{=}\,6$ for $\sim\!4\,{\times}\,10^4$ distributions. Importantly, we perform several rounds of analytical and numerical calculations to derive increasingly refined models for behaviors on the local boundaries. Details are given in appendix \ref{sec: appendix search}.

	\begin{figure}
		\includegraphics[width=\columnwidth]{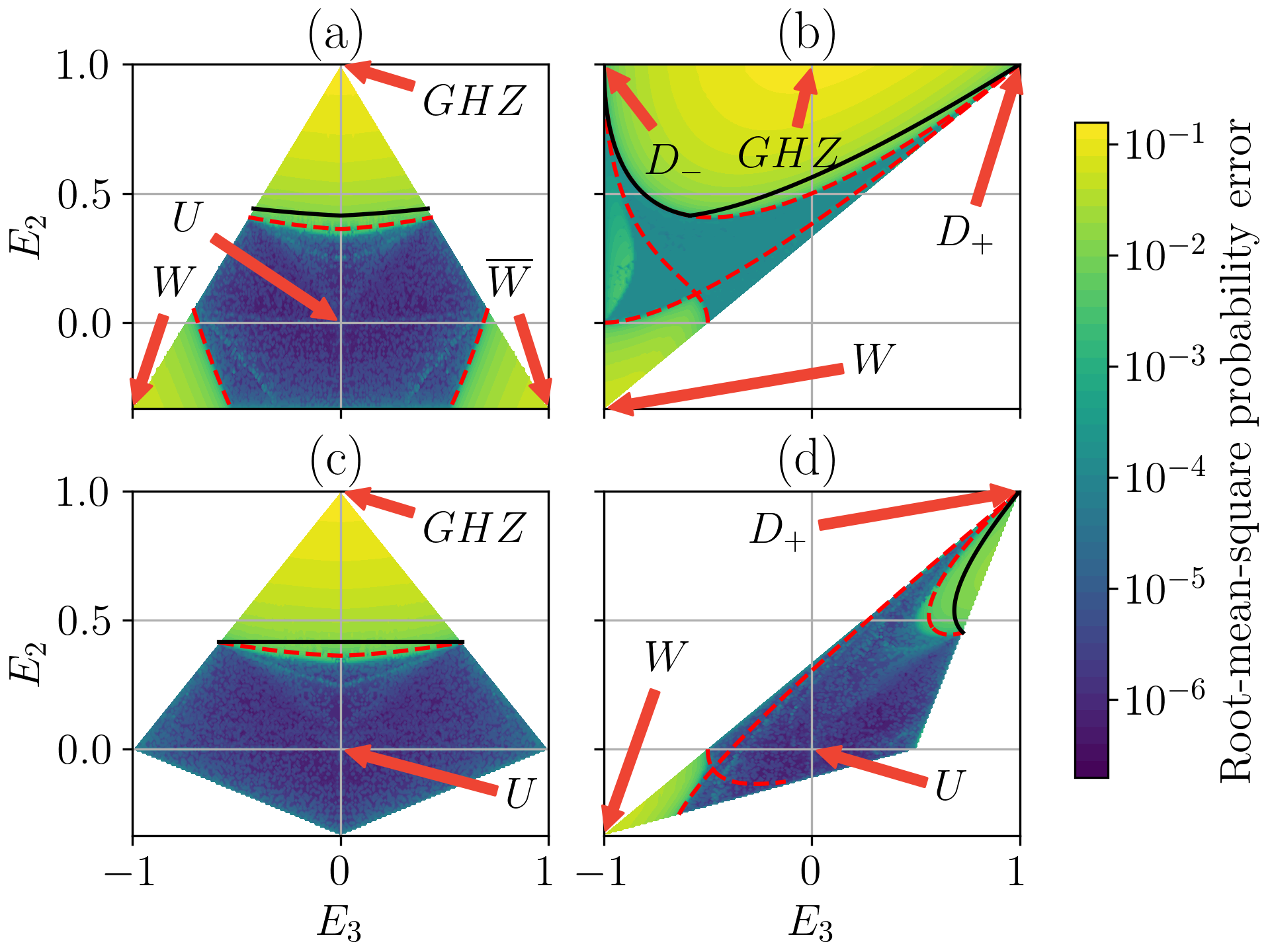}
		\caption{Euclidean distances between distributions of the form of Eq.~\eqref{eq:p} and best-fit triangle-local models \eqref{eq:local} in the following affine subspaces: (a) mixtures of $U$, $GHZ$ and $W$; (b) mixtures of $D_+$, $D_-$ and $W$; (c) the plane $E_1=0$; (d) mixtures of $U$, $D_+$ and $W$.
		The dashed lines are generated by our proposed boundary models appropriate for each subspace.
		The solid line is the no-signaling independence (NSI) boundary given by inequality (15) of Ref. \cite{NSI}.
		Regions with a potential for quantum non-locality are those between the dashed and solid lines.}
		\label{fig: 2d subspaces}
	\end{figure}

	Notably, while our search included the sufficient cardinalities of 666, all local models we found required smaller cardinalities. In fact, we find that all local distributions given by Eq.~\eqref{eq:p} can be described by strategies where the latent variable cardinalities are 333, 322, 422, or 332. These models are provided in appendix \ref{sec: appendix models}.

	In Fig.~\ref{fig: 2d subspaces} we show several sections of the tetrahedron of valid distributions that serve to illustrate our results.
	There, we plot the distance between the distribution and the best local model found.
	All panels show sharp transitions, which we use to calculate the analytical boundaries later on (in dashed red).

	Figure \hyperref[fig: 2d subspaces]{2(a)}  shows distributions in the plane $3E_1+E_3=0$, which contains as vertices different forms of ``full correlation'': the $GHZ$ distribution, corresponding to a shared random bit and characterized by $(E_1,E_2,E_3) = (0,1,0)$, the $W$ distribution, corresponding to the uniform mixture of the parties outputting such that $a+b+c=1$ and characterized by $(E_1,E_2,E_3) = (1/3,-1/3,-1)$, and the $\overline{W}$ distribution, which is defined as the $W$ distribution with all outcomes switched. It also contains the uniformly random distribution, $U$, characterized by $E_1=E_2=E_3=0$. This panel clearly shows the only three ``islands of nonlocality'' in the scenario [see also Fig.~\hyperref[fig:3d]{1(b)}].
	It also illustrates the symmetry $E_1,E_2,E_3\rightarrow -E_1,E_2,-E_3$ that is present, corresponding to switching all the output labels.

	Figure \hyperref[fig: 2d subspaces]{2(b)} depicts the plane $E_1+E_2-E_3=1$, which is spanned by combinations of the $W$ distribution, and the two only deterministic distributions in the scenario: $D_+$, described by $E_1=E_2=E_3=1$, and its outcome-relabeled version $D_-$.
	Here we find an interesting transition: For $E_3\leq-2+\sqrt{2}$ we find triangle-local models that saturate the inequality given in Ref.~\cite{NSI} for no-signaling and independence (NSI)-compliant distributions (depicted in black in the figure).
	However, for $E_3>-2+\sqrt{2}$, a gap between the NSI inequality and the triangle-local distributions appears, signaling a potential region for quantum nonlocality.

	Figure \hyperref[fig: 2d subspaces]{2(c)} illustrates the plane $E_1=0$, which was previously studied in Ref.~\cite{NSI} (see their Fig. 3).
	We reproduce their results, furthermore giving an analytic expression for the boundary (see \textit{Bell inequalities} below).

	Figure \hyperref[fig: 2d subspaces]{2(d)} contains the plane $E_1-2E_2+E_3=0$, which is spanned by $W$, $D_+$, and $U$.
	Here, another gap between the inequality in Ref.~\cite{NSI} and the distributions with triangle-local models appears, signaling another region of interest for settling the conjecture in Ref.~\cite{inflation_violations}.
	Interestingly, in the part of this figure close to the $W$ distribution, and also in the analogous region in Fig. \hyperref[fig: 2d subspaces]{2(b)}, the boundary showcases a kink.
	This is caused by the fact that, as we will see below, the boundary in this area is a piecewise composition of polynomial surfaces.
		
	\paragraph*{Bell inequalities.} As showcased above, only two boundaries need to be characterized, namely those around the $GHZ$ and $W$ distributions. The one close to $\overline{W}$ can be obtained from the $W$ boundary via the output relabeling transformation $E_1,E_2,E_3\rightarrow -E_1,E_2,-E_3$.

	In contrast to standard Bell scenarios, the non-convexity of the network-local set means a distribution must violate a set of inequalities, not just one, to be certified as nonlocal.
	This behavior is also seen in, e.g., full network nonlocality \cite{full,wang2023}.
	Therefore, in the context of networks, it is necessary to distinguish between Bell inequalities and tests of network nonlocality. In the minimal triangle network there are three archetypal forms of nonlocality: $GHZ$, $W$, and $\overline{W}$, each certified by a distinct test. Each test, in turn, consists of multiple inequalities.

\begin{figure}
		\includegraphics{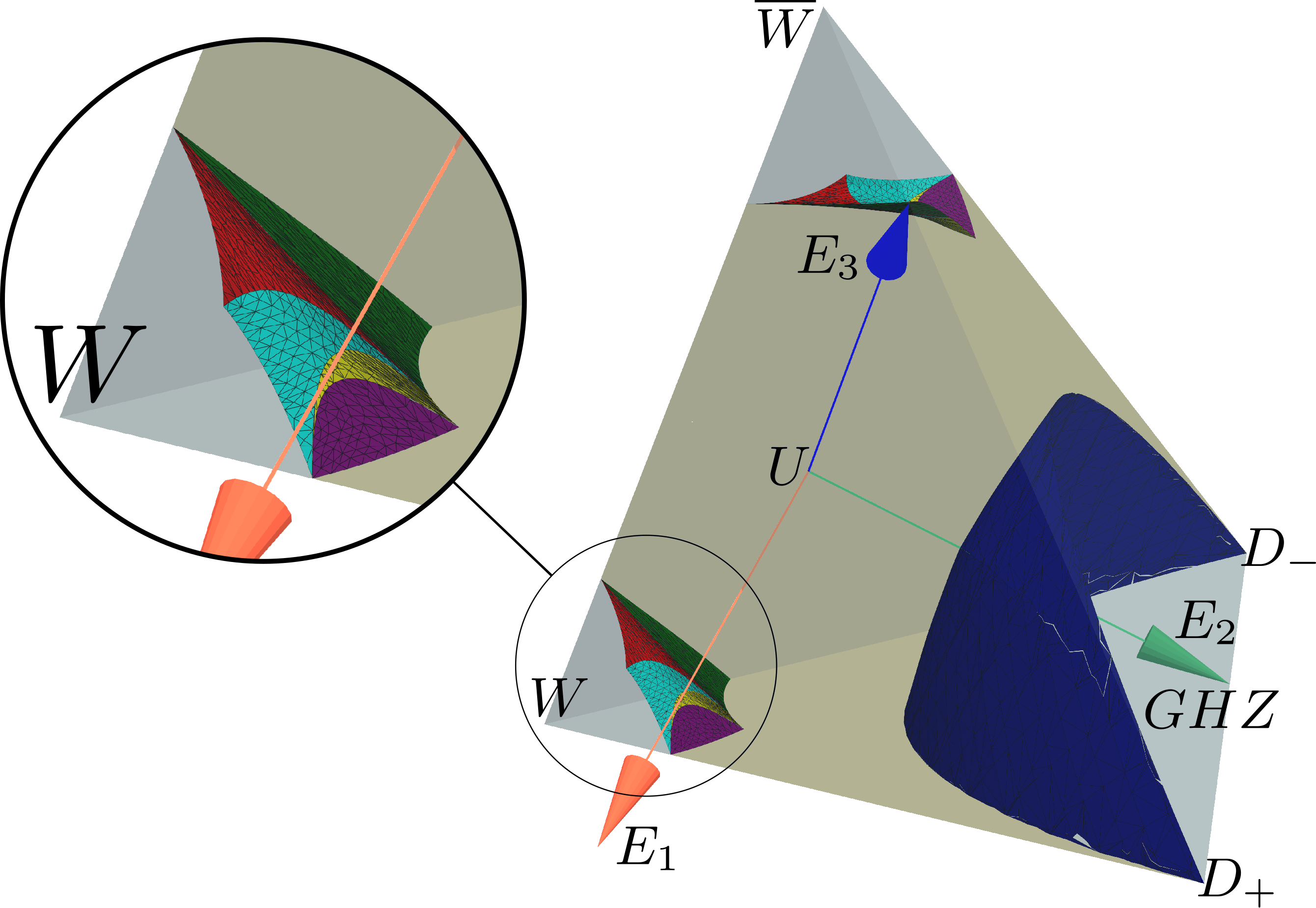}
		\caption{Close-up, rotated version of Fig.~\hyperref[fig:3d]{1(b)}. The $GHZ$ boundary (blue) is described by inequality (\ref{ineq GHZ}) and its outcome-relabeled version. The $W$ and $\overline{W}$ boundaries are each piecewise compositions of five distinct inequalities.
		The network-local models that generate the boundaries, and the inequalities can be found in appendices \ref{sec: appendix models} and \ref{sec: appendix inequalities}, respectively.}
		\label{fig:3d-zoom}
	\end{figure}

	The local boundaries are 2-dimensional surfaces, and therefore are spanned by families of models with two free parameters. The complete collection of boundary models can be found in appendix \ref{sec: appendix models}, but to give an example, consider the strategy with two degrees of freedom $(x,y)$ that generates the $GHZ$ boundary. The model is completely symmetric, having all hidden variables identically distributed, assuming values 1, 2, or 3, with probabilities $x$, $y$, and $1-x-y$, respectively. Moreover, all parties produce outputs deterministically according to the response function $f(\omega,\mu) = 1 - 2|\max(\omega,\mu)-2|$, so that the model is given by
	\begin{equation}
		\begin{aligned}
			&q(1)=r(1)=s(1)=x,\quad q(2)=r(2)=s(2)=y,\\
			&a = f(\beta,\gamma),\quad b = f(\gamma, \alpha),\quad c = f(\alpha,\beta).
		\end{aligned}
		\label{eq:GHZ model}
	\end{equation}
	From this model and its outcome relabeled version, we can calculate the three correlators and parameterize the $GHZ$ boundary with equations of the form $E_1 = E_1(x,y)$, $E_2 = E_2(x,y)$, and $E_3 = E_3(x,y)$. This surface is depicted in blue in Figs. \ref{fig:3d} and \ref{fig:3d-zoom}. Eliminating the variables $(x,y)$ from the parametric equations leads to the surface equation involving only correlators, as detailed in appendix~\ref{sec: appendix inequalities GHZ}. This expression for the boundary is then used in the $GHZ$ test. This gives that symmetric triangle-local distributions satisfy, at least, one of
	
	\begin{widetext}
		\begin{multline}
			8 (1 + E_1)^3 \left(11 + 37 E_1 + 28 E_1^2 - 4 E_1^3 + 8 E_2 + 13 E_1 E_2 + E_2^2 - 11 E_3 - 
			18 E_1 E_3 - 3 E_2 E_3 + 2 E_3^2\right) - (3 + 
			5 E_1 + E_2 - E_3)^4\\
			+ \sqrt{(3 + 5E_1 + E_2 - E_3)^2 - 8(1 + E_1)^3}\Big[\left(3+5E_1+E_2-E_3\right)^3-4\left(1+E_1\right)^3\left(7+11E_1+E_2-3E_3\right)\Big] \geq 0,\label{ineq GHZ}
		\end{multline}
	\end{widetext}
	and its equivalent by performing the transformation $E_1,E_2,E_3\rightarrow -E_1,E_2,-E_3$.

	When setting $E_1=0$, the resulting expression characterizes the network-local boundary found in Fig. 3 of Ref. \cite{NSI}, which was not explicitly calculated there. We provide this expression in appendix \ref{sec: appendix inequalities}. Moreover, as can be seen in Fig.~\hyperref[fig: 2d subspaces]{2(b)}, our boundary matches that obtained analytically in Ref.~\cite{NSI} for $E_3\leq-2+\sqrt{2}$. As shown in appendix \ref{sec: appendix nsi saturation}, Eq.~\eqref{ineq GHZ} recovers the inequality from Ref.~\cite{NSI} when restricted to this region, proving our result is tight therein.

	The strong numerical evidence that we build in this work leads us to conjecture that a criterion for detecting nonlocality in the binary-outcome triangle scenario is given by the simultaneous violation of Eq.~\eqref{ineq GHZ} and its outcome-relabeled version.
	Note that the square roots in these inequalities limit the distributions that can be tested with these witnesses, but if any of the expressions inside the square roots is negative, the behavior does not exhibit $GHZ$-type nonlocality.

	The boundaries close to $W$ and $\overline{W}$ are composed of five different regions (see Fig.~\ref{fig:3d-zoom}).
	Performing analogous variable elimination processes, aided by software for symbolic computation using Gr\"obner bases \cite{groebner_basis}, we give the remaining criteria for nonlocality, which consist of the simultaneous violations of five inequalities.
	We give such inequalities in appendix \ref{sec: appendix inequalities W}.

	\paragraph*{Perspectives for quantum nonlocality.} Whether the binary-outcome triangle network admits quantum nonlocality, a question first posed in Ref.~\cite{inflation_violations}, remains open despite significant effort (see, e.g., Refs. \cite{NSI,pozas2022proofs,pozas2023minimal,lauand2024,boreiri2023}).
	Our complete characterization of the set of symmetric triangle-local distributions allows us to provide insights on this question.
	Concretely, our work identifies three regions with the potential of supporting quantum nonlocality previously not considered in the literature, namely those in Figs. \hyperref[fig: 2d subspaces]{2(a)}, \hyperref[fig: 2d subspaces]{2(b)}, and \hyperref[fig: 2d subspaces]{2(d)} between the black lines (that represent the possible boundary of triangle-compatible distributions) and the red dashed lines that correspond to our inequalities.

	Moreover, we can revisit regions that were previously studied.
	In particular, we focus on the regions that appear in Figs. 2, 3 and Supplemental Fig. 2 in Ref.~\cite{NSI}.
	There, the regions colored in yellow correspond to distributions that do not violate their necessary conditions to be compatible with the triangle scenario, but their search for triangle-local models fails.
	Figure 3 in Ref.~\cite{NSI} corresponds to the area with $E_2\leq\frac12$ and $E_3\geq 0$ of Fig.~\hyperref[fig: 2d subspaces]{2(c)}.
	We find the same local boundary as the one reported in Ref.~\cite{NSI}, although we give an analytical expression for it, namely the particularization of Eq.~\eqref{ineq GHZ} provided in appendix \ref{sec: appendix inequalities GHZ}.

	\begin{figure}
		\centering
		\subfloat[\label{fig:e1e2}]{
			\begin{tikzpicture}[spy using outlines={circle, magnification=8, size=2cm, connect spies}]
				\node[anchor=south west, inner sep=0] (image) at (0,0) {\includegraphics[width=0.9\columnwidth]{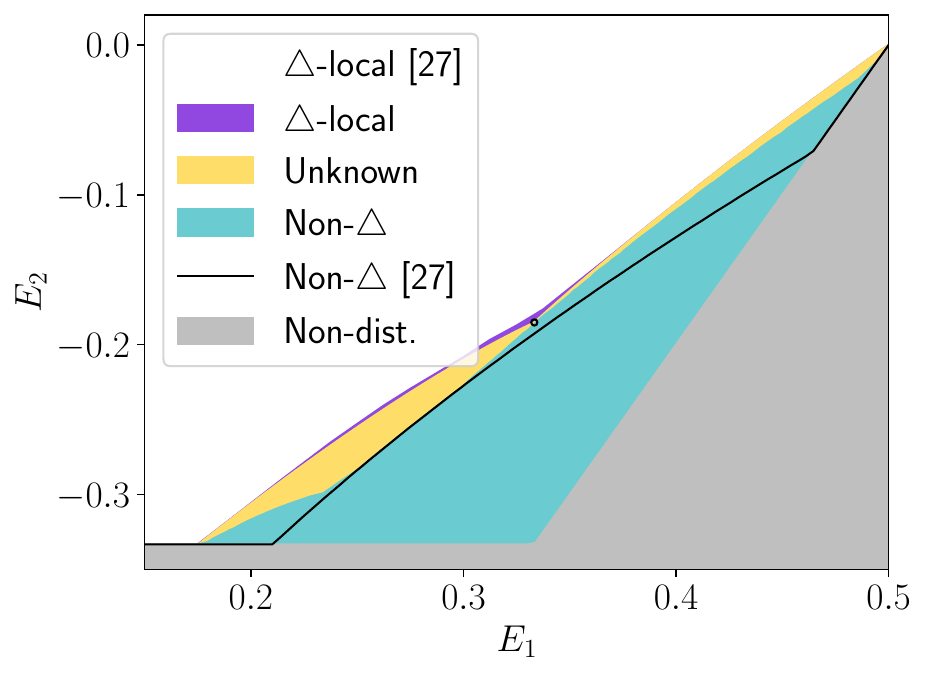}};
				
				\spy[black, size=1.8cm, magnification=5] on (4.5, 2.96) in node [fill=white] at (6.5, 1.9);
				\begin{scope}
					\draw[draw=black] (1.49,2.775) rectangle (2.15,3.005);
					\draw[draw=black] (1.49,2.775+0.91) rectangle (2.15,3.005+0.91);
					\draw[draw=black] (1.49,2.775+1.35) rectangle (2.15,3.005+1.35);
					\draw[draw=black] (1.49,2.775+1.79) rectangle (2.15,3.005+1.79);
					\draw[draw=black] (1.49,2.775+2.23) rectangle (2.15,3.005+2.23);
				\end{scope}
			\end{tikzpicture}
		}
		
		\subfloat[\label{fig:pq}]{
			\begin{tikzpicture}
				\node[anchor=south west, inner sep=0] (image) at (0,0) {\includegraphics[width=0.8\columnwidth]{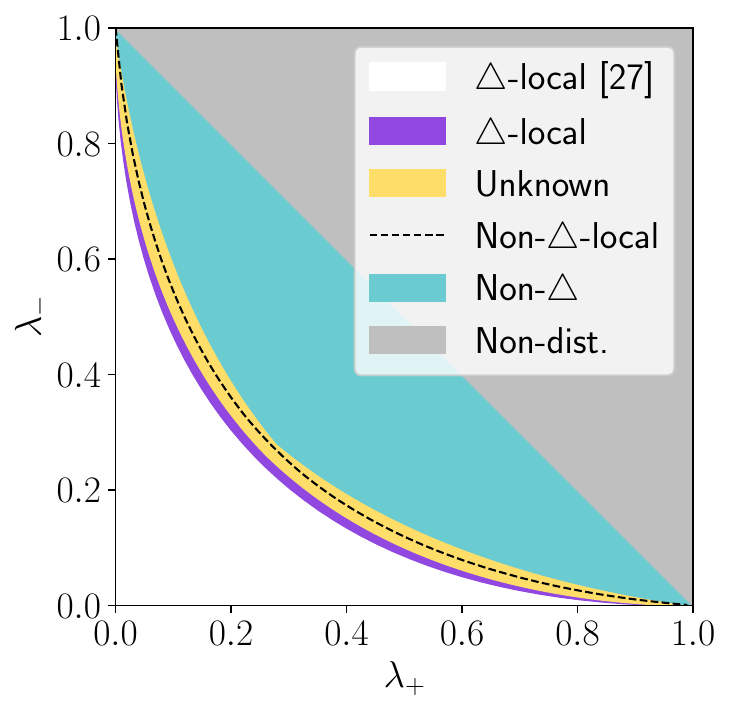}};
								
				\begin{scope}
					\draw[draw=black] (3.5,3.37) rectangle (4.222,3.65);
					\draw[draw=black] (3.5,3.37+0.5) rectangle (4.222,3.65+0.5);
					\draw[draw=black] (3.5,3.37+1.492) rectangle (4.222,3.65+1.492);
					\draw[draw=black] (3.5,3.37+1.985) rectangle (4.222,3.65+1.985);
					\draw[draw=black] (3.5,3.37+2.485) rectangle (4.222,3.65+2.485);
				\end{scope}
			\end{tikzpicture}}
		
		\caption{Reproductions of (a) Fig. 2 and (b) Supplemental Fig. 2 in Ref.~\cite{NSI}, with our improved characterizations. (a) depicts the projection of the behavior space onto the $E_1$-$E_2$ plane, while (b) depicts the plane containing the distributions $D_+$, $D_-$, and $\left(W + \overline{W}\right)/2$. The white regions contain the distributions that admit triangle-local models.
		The cyan regions contain the distributions that cannot be generated in the triangle network.
		The purple regions represent distributions for which we find triangle-local models, but were not identified in Ref.~\cite{NSI} to admit them.
		In (a), the solid line represents the limit to the blue region given in Ref.~\cite{NSI}.
		In (b), the dashed line represents our upper bound to the set of distributions admitting triangle-local models.
		Data for the white region in (a) were provided by the authors of Ref.~\cite{NSI} (private communication).}
		\label{fig:extend}
	\end{figure}

	For Fig. 2 and Supplemental Fig. 2 in Ref.~\cite{NSI}, we extend the triangle-local regions, thereby reducing the regions where quantum nonlocality could appear.
	We reproduce such figures in Fig.~\ref{fig:extend}, where the additional regions that we prove to have triangle-local models are depicted in purple.
	In Fig.~\hyperref[fig:e1e2]{4(a)}, a special distribution, corresponding to $(E_1,E_2,E_3)=(1/3,-5/27,-5/9)$, is marked by a black circle. Among the local distributions, this one exhibits the highest $\lambda_W$ in the parametrization $\lambda_W W + \lambda_+ D_+ + \lambda_- D_-+ (1-\lambda_W-\lambda_+-\lambda_-)\overline{W}$. It admits a model with cardinalities $c_\alpha=c_\beta=c_\gamma=3$, described by
	\begin{equation}\label{W example}
		\begin{aligned}
			&q(\alpha) = r(\beta) = s(\gamma) = \frac13\quad\forall\,\alpha,\beta,\gamma \in \{1,2,3\}, \\
			&A(-1|\beta,\gamma) = \delta_{\beta,1}\delta_{\gamma,1} + \delta_{\beta,1}\delta_{\gamma,3} + \delta_{\beta,2}\delta_{\gamma,3},\\
			&B(-1|\gamma,\alpha) = \delta_{\gamma,1}\delta_{\alpha,1} + \delta_{\gamma,2}\delta_{\alpha,1} + \delta_{\gamma,2}\delta_{\alpha,2},\\
			&C(-1|\alpha,\beta) = \delta_{\alpha,2}\delta_{\beta,3} + \delta_{\alpha,3}\delta_{\beta,2} + \delta_{\alpha,3}\delta_{\beta,3}.
		\end{aligned}
	\end{equation}
	
	In order to better analyze the potential for quantum nonlocality, we further constrain the set of distributions only subject to no-signaling and independence of the sources in the network, using inflation methods \cite{inflation}.
	The inflation corresponding to four copies of each of the sources (implemented with the \texttt{inflation} package in \textsc{Python} \cite{inflation_package}, see Ref. \cite{computationalappendix}) produces the cyan region in Fig.~\hyperref[fig:e1e2]{4(a)}, that improves over the boundary found in Ref.~\cite{NSI}.\footnote{One can also use inflation methods to approximate the set of triangle-local distributions, at an increased computational cost.
	For this reason, we only considered an inflation with three copies of one source and two copies of the remaining ones, which produced worse bounds than those obtained by considering inflations tailored for arbitrary triangle-compatible distributions.}
	In combination, our explicit triangle-local constructions and the improved outer approximations provided by inflation significantly reduce the set of distributions that could, in principle, demonstrate quantum nonlocality.
	We expect that improved characterizations of the theory-agnostic distributions compatible with the triangle scenario in this region will reduce it further until closing the gap.

	In Fig.~\hyperref[fig:pq]{4(b)} we also show that the region of distributions that admit triangle-local models is larger than that reported in Ref.~\cite{NSI}.
	Our boundary corresponds to the inequality \eqref{ineq GHZ}, that we build from models where all the cardinalities are three.\footnote{The boundary given in Ref.~\cite{NSI} is recovered by searching for models where one hidden variable has cardinality 3 while the remaining ones have cardinality 2. Interestingly, the model we find in our improved region appears in Ref.~\cite{NSI}, namely in Eq. (24) in their Supplemental Material, but is not used for creating Supplemental Fig. 2 there.}
	Here, inflation methods for triangle-compatible distributions did not produce improvements, suggesting that the inequality given in Ref.~\cite{NSI} may be tight in this region.
	However, using inflation tailored for triangle-local distributions we obtain an outer approximation [the black dashed line in Fig.~\hyperref[fig:pq]{4(b)}] that approaches closely our conjectured set.
	We expect that the remaining gap will be closed by improved outer characterizations.
	These results, despite reducing the region that could potentially host quantum nonlocality via the inner constructions, still signal it as a promising candidate where further effort should be put. 
	
	\paragraph*{Discussion.} In analogy with the study of Bell nonlocality, characterizing the probability distributions generated in networks that can be reproduced within classical physics is crucial for identifying genuinely quantum phenomena that can be exploited in applications such as quantum cryptography.
	In this work, using a combination of strong numerical analysis and analytical calculations, we give what we conjecture to be the complete characterization of the set of distributions symmetric under permutations of parties that admit a network-local model in the binary-outcome triangle scenario.

	The analysis, based on both explicit constructions of network-local models and outer approximations of the whole set, was performed using readily available tools \cite{localmodels,inflation_package} that can be easily adapted to more complex networks and input-output scenarios.
	These tools, or improved ones, will have a prominent role in the analysis of quantum networks. 
	However, the fact that the amount of resources scales unfavorably with the complexity of the scenario motivates the search for more efficient characterization methods.

	We show that symmetric, binary-outcome, triangle-local distributions can be characterized without saturating the cardinality bounds of Ref.~\cite{cardinality_bound}.
	Why this is so, and whether non-symmetric distributions saturate the bound, remain interesting open questions.
	
	Our improved characterization, while based primarily on numerical methods, inequivocally rules out some regions of the parameter space thought to be promising for demonstrating quantum nonlocality. Other regions (including new ones that we identify), however, persist as valid candidates to support genuinely quantum behaviors.
	Still, more work is needed to decide the conjecture of Ref.~\cite{inflation_violations}, which crucially will require improved methods for explicitly constructing quantum realizations in networks.

	\paragraph*{Acknowledgments.} This work received financial support from the Brazilian agencies Coordena\c{c}\~ao de Aperfei\c{c}oamento de Pessoal de N\'{\i}vel Superior (CAPES), Funda\c{c}\~ao de Amparo \`a Ci\^encia e Tecnologia do Estado de Pernambuco (FACEPE - Grant No. BPP-0037-1), Conselho Nacional de Desenvolvimento Cient\'{\i}fico  e Tecnol\'ogico through its program CNPq INCT-IQ (Grant No. 465469/2014-0), Funda\c{c}\~ao de Amparo \`a Pesquisa do Estado de S\~ao Paulo (FAPESP - Grant No. 2021/06535-0), and the Swiss National Science Foundation (Grant No. 224561). Computations were performed in part at University of Geneva using the Baobab HPC service.

	\paragraph*{Data availability.} The data that support the findings of this article are openly available \cite{computationalappendix}.
	
	\bibliography{references.bib}
	
	\onecolumngrid
	
	\appendix
	
	\section{Details on the numerical search}
	\label{sec: appendix search}
	
	In this appendix we provide all the details of the determination of the set of triangle-local distributions and of its boundaries. The workflow is summarized in Fig.~\ref{fig:model_search}. Each step is described below in detail.

\begin{figure}[h]
    \centering
    \begin{tikzpicture}[node distance=2cm]

        \node (start) [startstop] {Start};

        \node (explore) [numerical, below of=start, align=center] {Exploratory search for \\ models in 2d subspaces};

        \node (inference) [analytical, below of=explore, align=center] {Model inferences for points \\ close to local boundary \\ and surface extraction};

        \node (surface) [numerical, below of=inference, align=center] {Boundary validation \\ procedure};

        \node (violations) [decision, below of=surface, yshift=-1cm, align=center] {Are there any \\ violations?};

        \node (end) [startstop, below of=violations, yshift=-1.5cm] {End};

        \node (data) [decision, right of=violations, xshift=3cm, align=center] {Is there enough \\ data to try model \\ inference?};

        \node (add) [analytical, above of=data, yshift=5cm, align=center] {Add more 2d \\ subspaces};

        \draw [arrow] (start) -- (explore);
        \draw [arrow] (explore) -- (inference);
        \draw [arrow] (inference) -- (surface);
        \draw [arrow] (surface) -- (violations);

        \draw [arrow] (violations) -- node[anchor=east] {No} (end);
        \draw [arrow] (violations.east) -- node[anchor=south] {Yes} (data.west);

        \draw [arrow] (data.north) |- node[anchor=west] {Yes} (inference.east);
        \draw [arrow] (add.west) -- (explore.east);

        \draw [arrow] (data.east) -- ++(1.5,0) node[anchor=north] {No} |- (add.east);

        \node [numerical, right of=add, xshift=4.5cm, minimum width=2.5cm] (legend_num) {Numerical Step};
        \node [analytical, below of=legend_num, yshift=0.5cm, minimum width=2.5cm] (legend_ana) {Analytical Step};

    \end{tikzpicture}
    \caption{Flowchart of the search for triangle-local models in this work. Numerical steps are highlighted in blue, and analytical steps are highlighted in green.}
    \label{fig:model_search}
\end{figure}

	Our process begins with the selection of some 2-dimensional affine subspaces in the behavior space, whose distributions are subjected to an exploratory search of local models. The subspaces that were investigated in the first round of the analysis were the affine combinations of i) $U$, $GHZ$, and $W$, ii) $D_+$, $D_-$, and $W$, iii) $U$, $D_+$, and $W$, iv) $U$, $D_+$, and $GHZ$, as well as v) the plane $E_1 = 0$. All of them contain both local distributions (such as $U$, $D_+$, $D_-$) and nonlocal ones (such as $GHZ$, $W$, $\overline{W}$), so they all feature local-nonlocal transitions, as can be seen in Fig.~\ref{fig: 2d subspaces}. Moreover, the plane $E_1 = 0$ has been previously analyzed in Ref. \cite{NSI}.

	In each plane, we sample approximately 10,000 distributions by uniformly sampling $E_1$, $E_2$, and $E_3$ subject to the corresponding constraints. For each of these, we try to fit a local model to the target distribution, by minimizing the sum of squared errors on the probabilities. This is a non-convex optimization problem over $(c_\alpha - 1) + (c_\beta - 1) + (c_\gamma - 1) + c_\beta c_\gamma + c_\gamma c_\alpha + c_\alpha c_\beta$ variables, where $c_\alpha$, $c_\beta$, and $c_\gamma$ are the hidden variable cardinalities. We implement and solve this problem using the \texttt{scipy} library in \textsc{Python} (see the computational appendix \cite{computationalappendix}). Initially, we set the hidden variable cardinalities to $c_\alpha=c_\beta=c_\gamma=6$, which is guaranteed to generate the complete triangle-local set \cite{cardinality_bound}. For subsequent rounds, the computations were expedited by lowering the cardinalities to $c_\alpha = c_\beta = c_\gamma = 4$, or even $c_\alpha = c_\beta = c_\gamma = 3$ in some subspaces, since these reduced cardinalities were enough to generate points on the local boundary. For each distribution, we run the optimization 200 times, each one with a different, uniformly random initial guess for $q$, $r$, $s$, $A$, $B$, and $C$, and take the model with the lowest error to the target behavior.

	After the search for models, we look at those with root-mean-square probability error of the order of $10^{-4}$, since these are close to the sharp transitions of the error heat maps, such as the ones depicted in Fig.~\ref{fig: 2d subspaces}. These models typically have several parameters that take on values close to $0$ or $1$. The model inference procedure then consists of fixing these parameters to be exactly equal to the extreme values, and solving for the remaining parameters by enforcing the symmetry constraints. Unlike the exploratory analysis, this is done analytically (see a more detailed discussion in appendices \ref{sec: appendix inequalities} and \ref{sec: appendix optimization}).
	
	The output of the model inference step are models which turn out to have two free parameters, thus they span surfaces in $(E_1, E_2, E_3)$ space. These surfaces are joined to form the tentative boundaries of the local set, which perform well in capturing the local-nonlocal transitions in the 2-dimensional error plots. Naturally, this is expected, since the boundary models are derived from the optimization results in the very same regions, so a more meaningful test for these surfaces should consider the whole behavior space, going beyond the affine subspaces of the exploratory analysis. That is the purpose of the boundary validation procedure, following the flowchart of Fig.~\ref{fig:model_search}.

	For this test, we sample 1,000 distributions on each of the proposed boundaries, and slightly displace them by an euclidean distance of $10^{-3}$ (in correlator space) towards the archetypal nonlocal distributions, $GHZ$ or $W$ depending on the boundary being tested, so that we have distributions just outside the provisional local set. These 2,000 distributions (the original 1,000, and the corresponding displaced ones) are then subjected to a local model search with 1.000 trials per distribution. The optimization runs of this boundary validation procedure are always performed with the full cardinalities $c_\alpha = c_\beta = c_\gamma = 6$.

	Out of these 2,000 optimized models, the ones that achieve probability error of the order of $10^{-4}$ or lower are further analyzed. Most of them are equivalent to the boundary models, when considering hidden variable exchanges and relabellings. The genuinely new ones generate points that, despite being beyond the proposed local boundary, admit a triangle-local model. These are the violations mentioned in Fig.~\ref{fig:model_search}, that indicate that the local boundary should be improved with more powerful models. While the boundary close to the $GHZ$ distribution had no violations in the validation test already in the first run, the boundary close to the $W$ distribution needed three cycles of gradual improvement in order to finally pass this check.

	If the general structure of the improved models can be inferred from the boundary violations, the new cycle begins with the model inference step. Otherwise, we add more subspaces to the exploratory analysis phase, as shown in Fig.~\ref{fig:model_search}. These additional subspaces are chosen in a way to be close to the violations found in the previous step, so as to capture the features that were missing in the previous set of subspaces. At the end of the process, we had to add the following planes to our initial collection: vi) $3E_1 + 3E_2 - E_3 = 1$, vii) $E_1 + (4+2\sqrt{5})E_2 + E_3 = 0$, viii) $12E_1 - 4E_2 + 6E_3 = 1$, as well as ix) the affine combinations of $D_+$, $W$, and $\overline{W}$.

	\section{Models for distributions at the boundaries}
	\label{sec: appendix models}
	
	In this appendix we give the concrete families of triangle-local models, defined by Eq.~\eqref{eq:local}, that describe the distributions invariant under permutations of parties that lie in the boundaries of the network-local set for the binary-outcome triangle network.
	For ease of notation, we represent $q$, $r$ and $s$ by one-dimensional arrays that contain the corresponding probabilities, and $A\equiv A(1|\beta,\gamma$), $B\equiv B(1|\gamma,\alpha)$ and $C\equiv C(1|\alpha,\beta)$ by two-dimensional arrays.
	In each of these, the first hidden variable indexes the rows, while the second one indexes the columns.
	The arrays corresponding to the outcomes $-1$ can be obtained by using the normalization condition.
	
	\subsection{Models for the boundary close to the GHZ distribution}
	The boundary close to the $GHZ$ distribution (in blue in Figs.~\ref{fig:3d} and \ref{fig:3d-zoom}) is described by a single family of models.
	As described in Eq.~\eqref{eq:GHZ model}, it has cardinalities $c_\alpha=c_\beta=c_\gamma=3$ and is given by
	\begin{equation}\label{GHZ example 1}
		q(\omega) = r(\omega) = s(\omega) = \begin{cases} x & \omega=1 \\ y & \omega=2 \\ 1-x-y & \omega=3 \end{cases}, \qquad\begin{matrix} A(1|\omega,\mu) \\ B(1|\omega,\mu) \\ C(1|\omega,\mu) \end{matrix} = \begin{cases} 1 & (\omega,\mu)\in\{(1,2),(2,1),(2,2)\}\\ 0 & \text{otherwise} \end{cases}
	\end{equation}
	for arbitrary, nonnegative $x$ and $y$ satisfying $x+y\leq1$.
	In the notation that was described above, this model takes the form
	\begin{align}
		&q = r = s = \begin{bmatrix}x & y & 1-x-y\end{bmatrix},\,A = B = C = \begin{bmatrix}0 & 1 & 0\\
			1 & 1 & 0 \\
			0 & 0 & 0\end{bmatrix}.
			\label{GHZ}
	\end{align}
	This model characterizes only half of the boundary.
	The other half is described by the model with the outcomes flipped, which amounts to doing $A\rightarrow 1-A$, $B\rightarrow 1-B$ and $C\rightarrow 1-C$ in Eq.~\eqref{GHZ}.
	Note that the family is described by two free parameters, $x$ and $y$.
	
	\subsection{Models for the boundary close to the W distribution}
	
	The situation is considerably richer for the boundary close to the $W$ distribution, where there are five different classes of models that generate different sections of the surface, as depicted in Fig.~\ref{fig:3d-zoom}.
	The models that characterize the boundary close to the $\overline{W}$ distribution are the same as those presented below when replacing $A\rightarrow 1-A$, $B\rightarrow 1-B$ and $C\rightarrow 1-C$.

	\subsubsection{Model for the green surface}
	The distributions on the green surface admit the following model with cardinalities $c_\alpha=3$, $c_\beta=c_\gamma=2$:
	\begin{align}
		&q = \begin{bmatrix}1-2x & x& x\end{bmatrix}, \,
		r = s = \begin{bmatrix}y &  1-y\end{bmatrix},\, A = \begin{bmatrix}z(x,y) & 1\\
			1 & t(x,y)\end{bmatrix},\,B = \begin{bmatrix}1& 0 & 1\\
			0& 0 & 1\end{bmatrix},\,C = \begin{bmatrix}1 & 0\\
			1 & 1\\
			0 & 0\end{bmatrix},\label{W1}
	\end{align}
	where $z(x,y)=\dfrac{2 x^2 y - x^2 -4 x y^2 + x y + 2 y^2 - y}{y^2(1-2x)}$ and $t(x,y)=\dfrac{x(1 - x - y + 2 x y)}{1 - 2x - 2y + 4xy + y^2 - 2xy^2}$.
	The free parameters of the family are $x$ and $y$, which satisfy $0\leq x \leq x_\mathrm{max}$ and $1+ x + \sqrt{\dfrac{1 + 5x^2 - 2x^3}{1 - 2x}} \leq 4y\leq 4 - 2x\left(1 + \sqrt{\dfrac{5-2x}{1 - 2x}}\right),$ $x_\mathrm{max}$ being the smallest real root of the polynomial $9 x^4 - 24 x^3 + 24 x^2 - 9 x + 1$, namely
	\begin{equation*}
		\begin{aligned}
			x_\mathrm{max}=&\,\frac{2}{3}-\frac{1}{6 \sqrt{\frac{3}{\sqrt[3]{\frac{675}{2}-\frac{81 \sqrt{41}}{2}}+3 \sqrt[3]{\frac{1}{2} \left(3 \sqrt{41}+25\right)}}}}\\
			&-\frac{1}{2} \sqrt{-\frac{1}{27} \sqrt[3]{\frac{675}{2}-\frac{81 \sqrt{41}}{2}}-\frac{1}{9} \sqrt[3]{\frac{1}{2} \left(3 \sqrt{41}+25\right)}+\frac{10}{3 \sqrt{3 \left(\sqrt[3]{\frac{675}{2}-\frac{81 \sqrt{41}}{2}}+3 \sqrt[3]{\frac{1}{2} \left(3 \sqrt{41}+25\right)}\right)}}}\\
			\approx&\,0.1916.
		\end{aligned}
	\end{equation*}
	
	\subsubsection{Model for the purple surface}
	The distributions on the purple surface admit the following model with cardinalities $c_\alpha=4$, $c_\beta=c_\gamma=2$:
	\begin{align}
		&q = \begin{bmatrix}z(x,y) & z(x,y) & t(x,y) & 1-2z(x,y)-t(x,y)\end{bmatrix}, \,
		r = s = \begin{bmatrix}x &  1-x\end{bmatrix},\,A= \begin{bmatrix}0 & 1\\
			1 & 1\end{bmatrix},\,B = \begin{bmatrix}
			y& 1 & 1 & 0\\
			0& 1 & 0 & 1
		\end{bmatrix},\,C = \begin{bmatrix}
			1 & 1\\
			y & 0\\
			1 & 0\\
			0 & 1
		\end{bmatrix},\label{W2}
	\end{align}
	where $z(x,y) = \dfrac{(x-1)^3x}{1 - 4 x + 6 x^2 - 4 x^3 - x y + x^2 y + x^3 y}$, $t(x,y) = \dfrac{(x-1)x^3(1-y)}{1 - 4 x + 6 x^2 - 4 x^3 - x y + x^2 y + x^3 y}$.
	Again, $x$ and $y$ are the free parameters of the family of models.
	In this case, they satisfy $0\leq y \leq 1$ and $x_\mathrm{min}\leq x \leq 1$, $x_\mathrm{min}$ being the only positive root of the polynomial $(3-y)x^4 - 3x^3 - y x^2 + (2+y)x - 1$ for $y\not=1$ and $x_\mathrm{min}(y=1)=\frac12$.
	
	\subsubsection{Model for the red surface}
	The distributions on the red surface admit the following model with cardinalities $c_\alpha=c_\beta=c_\gamma=3$:
	\begin{align}
		&q = r=\begin{bmatrix}x & z(x,y)& 1-x-z(x,y)\end{bmatrix}, \,
		s = \begin{bmatrix}y &  \dfrac{1-y}{2} & \dfrac{1-y}{2}\end{bmatrix},\notag\\
		&A = \begin{bmatrix}
			0 & 1 & 0\\
			1 & 1 & 0\\
			1 & 1 & 1
		\end{bmatrix},\,B = \begin{bmatrix}
			0 & 1 & 1\\
			0 & 0 & 1\\
			1 & 1 & 1\end{bmatrix},\,C = \begin{bmatrix}
			t(x,y) & 1 & 1\\
			1 & u(x,y) & 0\\
			1 & 0 & 0\end{bmatrix},\label{W3}
	\end{align}
	where the functions $z$, $t$, and $u$ are $z(x,y) = \dfrac{x - x^2 - 2 y + 4 x y - x^2 y + x y^2}{(x-y)(1-y)}$, $t(x,y) = \dfrac{x - 4 (1 - x)^3 y + (3 - 4 x) x y^2}{4x^2y(x-y)}$, and $u(x,y) = \dfrac{(x - y) (1 - y)^2 [4 y (1 + x^2 + x^3 - x^2 y) - x  -x y (10+y)]}{4 y (x - x^2 - 2 y + 4 x y - x^2 y + x y^2)^2}$.
	As before, the probabilities $x$ and $y$ are the free parameters of the family, which satisfy $1/3 \leq y \leq 1$ and $ \frac{1}{24y}\mathrm{max}\Big[1+12y+3y^2-\sqrt{\left(1+12y+3y^2\right)^2-192y^2},$
	$12y\left(2+y-\sqrt{2+2y+y^2}\right)\Big]\leq x \leq x_\mathrm{max}$, $x_\mathrm{max}$ being the smallest root of the polynomial $16 y^2 x^4 -12 y (1 + 2 y + 5 y^2) x^3 + (1 + 12 y + 34 y^2 + 84 y^3 + 45 y^4) x^2 -(5 + 16 y + 50 y^2 + 24 y^3 + y^4) y x + 4y^2(1+y)^2$.
	
	The model given in Eq. \eqref{W example} produces a behavior on the intersection of the red and cyan surfaces in Fig.~\ref{fig:3d-zoom}, so it can be regarded as a member of the family \eqref{W3}, with $x=\frac{1}{3} + \varepsilon$, $y = \frac{1}{3} + 4\varepsilon$ and then taking the limit $\varepsilon\rightarrow0$.
	
	\subsubsection{Model for the yellow surface}
	The distributions on the yellow surface admit the following model with cardinalities $c_\alpha=c_\beta=3$, $c_\gamma=2$:
	\begin{align}
		&q = \begin{bmatrix}
			z(x,y) & t(x,y) & 1-z(x,y)-t(x,y)\end{bmatrix}, \notag \\
		&r = \begin{bmatrix}u(x,y) &  v(x,y) & 1-u(x,y)-v(x,y)\end{bmatrix}, \notag \\
		&s = \begin{bmatrix}
			x & 1-x
		\end{bmatrix}, \notag \\
		&A = \begin{bmatrix}
			0 & 1\\
			1 & 1\\
			0 & 0
		\end{bmatrix},\,B = \begin{bmatrix}
			y & 1 & 1\\
			1 & 1 & 0\end{bmatrix},\,C = \begin{bmatrix}
			0 & 0 & 1\\
			1 & 0 & 1\\
			1 & 1 & 1\end{bmatrix}.\label{W4}
	\end{align}
	In the model above, $z(x,y)$ is the largest root of the polynomial
	\begin{equation}
		\begin{aligned}
			(1 &- y)^2 (1 - x + x^2 y) (1 - x + x y + x y^2)z^2 \\
			&- (1 - y) (x - 2 x^2 + x^3 - 2 y + 5 x y - 3 x^2 y + x^3 y - x^4 y - x^2 y^2 + x^3 y^2 + x^4 y^2 + x^2 y^3 - x^3 y^3)z \\
			&- (1 - x) (x - y) y (1 - x + x y).
		\end{aligned}\nonumber
	\end{equation}
	Based on it, and on the free parameters $x$ and $y$ (constrained only by the requirements of all probabilities being positive), the remaining parameters are
	\begin{align}
		t(x,y) &= \dfrac{1 - 2 z - x + 2 x z - x y^2 z}{1-x+y},\notag\\
		u(x,y) &= \dfrac{(1-y)(1 - z - x + x z + y z + x y - x y z - x^2 y z + x^2 y^2 z)}{(1-x+y)(1 - x + x y)},\notag\\
		v(x,y) &= \frac{y}{1-y}\, u(x,y).\notag
	\end{align}
	
	\subsubsection{Model for the cyan surface}
	Finally, distributions on the cyan surface admit the following model with cardinalities $c_\alpha=c_\beta=c_\gamma=3$:
	\begin{align}
		&q = \begin{bmatrix}
			x & y & 1-x-y\end{bmatrix}, \, 
		r = \begin{bmatrix}
			z & t & 1-z-t
		\end{bmatrix},\,
		s = \begin{bmatrix}u &  v & 1-u-v\end{bmatrix},\,A = \begin{bmatrix}
			0 & 1 & 0\\
			1 & 1 & 0\\
			1 & 1 & 1
		\end{bmatrix},\,B = \begin{bmatrix}
			0 & 1 & 1\\
			0 & 0 & 1\\
			1 & 1 & 1\end{bmatrix},\,C = \begin{bmatrix}
			w & 1 & 1\\
			1 & 1 & 0\\
			1 & 0 & 0\end{bmatrix},\label{W5}
	\end{align}
	where the seven probabilities $x$, $y$, $z$, $t$, $u$, $v$, and $w$ can be parameterized as complicated functions of the correlators $E_1$ and $E_2$, that serve as the two free parameters for the model. The complete characterization of this model is provided in appendix \ref{sec: appendix optimization}. 
	
	As mentioned before, the model described in Eq. \eqref{W example} is also a member of the family \eqref{W5}. By fixing $E_1 = 1/3$ and $E_2 = -5/27$, and following the procedure of appendix \ref{sec: appendix optimization}, the parameters $x$ and $z$ are obtained exactly by solving a system of polynomial equations that has 44 solutions, one of them being $x=z=1/3$. The other parameters readily follow from the symmetry and correlator constraints. Alternatively, the same result can be obtained numerically with the routine that generates points on the cyan surface, provided in Ref.~\cite{computationalappendix}.
	
	\section{List of tests for triangle nonlocality}
	\label{sec: appendix inequalities}
	In this appendix we give both the Bell inequalities that characterize the boundaries of the set of distributions that admit triangle-local models, and the associated tests of nonlocality.
	As explained in the main text, the direct connection between Bell inequalities and tests for nonlocality that is drawn in Bell-type scenarios cannot be made in networks due to the fact that the sets of network-local correlations are not convex.
	A distribution can be considered as network-nonlocal only if it violates, in general, several Bell-like inequalities.
	Thus, we distinguish between Bell inequalities (i.e., the boundaries of the set of distributions that admit network-local models), and tests for network nonlocality.

	Each of the subsections below describes a different test of network nonlocality.
	Passing any of them identifies a distribution as network-nonlocal.
	However, each of the tests involves the violation of several inequalities, and all the inequalities in a test must be violated in order to pass the test.
	The inequalities that determine the boundary of the set of symmetric, network-local distributions are Eqs.~\eqref{ineq GHZ repeat}, \eqref{ineq W1}, \eqref{ineq W2}, \eqref{ineq W3}, \eqref{ineq W4}, and \eqref{ineq W5}, as well as their counterparts under the transformation $E_1,E_2,E_3\rightarrow -E_1,E_2,-E_3$.

	The computational appendix \cite{computationalappendix} for this work contains \textsc{Python} implementations of all the tests and inequalities described below, routines to generate points on the local boundaries, and a script to render an interactive version of Fig.~\hyperref[fig:3d]{1(b)}.

	\subsection{Test for network nonlocality close to the GHZ distribution}
	\label{sec: appendix inequalities GHZ}
	The family of models that describe the boundary close to the $GHZ$ distribution is given by Eq.~\eqref{GHZ}. In terms of its degrees of freedom $(x,y)$, this model has the following correlators:
	\begin{subequations}
		\begin{align}
			E_1 &= -1 + 4 x y + 2y^2,\label{E1 GHZ}\\
			E_2 &= 1 - 8 x y + 4 x^2 y - 4 y^2 + 12 x y^2 + 4 y^3,\label{E2 GHZ}\\
			E_3 &= -1 + 12 x y - 12 x^2 y + 6 y^2 - 12 x y^2 - 4 y^3.\label{E3 GHZ}
		\end{align}
	\end{subequations}
	Equation~\eqref{E1 GHZ} can easily be solved for $x$, giving $x = (1+ E_1 - 2y^2)/(4y)$.
	Substituting this into Eqs. \eqref{E2 GHZ} and \eqref{E3 GHZ}, we get depressed quartic equations in $y$, namely
	\begin{subequations}
	\begin{align}
		E_2 &= \frac{(1+E_1)^2}{4y} - 1 - 2E_1 + 2y(1+E_1) - y^3, \label{eq:E2sub}\\
		E_3 &= -\frac{3(1+E_1)^2}{4y} + 2 + 3E_1 - y^3. \label{eq:E3sub}
	\end{align}
	\end{subequations}
	Subtracting Eq.~\eqref{eq:E3sub} from Eq.~\eqref{eq:E2sub} yields a simpler quadratic equation.
	Solving it provides an expression for $y$ in terms of the correlators, ultimately resulting in a single equation that leads to the inequality
	\begin{multline}
		8 (1 + E_1)^3 \left(11 + 37 E_1 + 28 E_1^2 - 4 E_1^3 + 8 E_2 + 13 E_1 E_2 + E_2^2 - 11 E_3 - 
		18 E_1 E_3 - 3 E_2 E_3 + 2 E_3^2\right) - (3 + 
		5 E_1 + E_2 - E_3)^4\\
		+ \sqrt{(3 + 5E_1 + E_2 - E_3)^2 - 8(1 + E_1)^3}\Big[\left(3+5E_1+E_2-E_3\right)^3-4\left(1+E_1\right)^3\left(7+11E_1+E_2-3E_3\right)\Big] \geq 0. \label{ineq GHZ repeat}
	\end{multline}
	Proceeding analogously with the outcome-relabeled version of the model \eqref{GHZ} (which amounts to performing the transformation $E_1\rightarrow -E_1$, $E_2\rightarrow E_2$, $E_3\rightarrow -E_3$), we obtain the second inequality that characterizes the boundary of the network-local close to the $GHZ$ distribution.

	The associated test of nonlocality consists of the simultaneous violation of both Eq.~\eqref{ineq GHZ repeat} and its outcome-relabeled version.
	In other words, a distribution is guaranteed not to admit a triangle-local model if it violates, both, Eq.~\eqref{ineq GHZ repeat} and its outcome-relabeled version.
	
	Note that the inequalities contain square roots which could, in principle, take negative values for some distributions.
	For instance, the argument of the square root in Eq.~\eqref{ineq GHZ repeat} evaluates to $-\frac{64}{27}$ in the $\overline{W}$ distribution.
	This sets a limit on the distributions that can be detected with this test to only those for which (naturally) the left-hand sides evaluate to negative real numbers.
	When any of the left-hand sides is positive or complex, either the corresponding distribution admits a triangle-local model or it violates one of the remaining tests, that we give in the next section.

	To conclude the explanation of the $GHZ$ test, we provide the analytical characterization of the local boundary for the subspace $E_1=0$, shown in Fig.~\hyperref[fig: 2d subspaces]{2(c)}. As mentioned in the main text, the local region for this subspace is also depicted in Fig. 3 of Ref. \cite{NSI}, but a characterization in terms of the correlators is not provided. This can be found by simply substituting $E_1 = 0$ in inequality (\ref{ineq GHZ repeat}) and its outcome-relabeled version. If we introduce a new variable $u = (3 + E_2 - |E_3|)/2$, it is straightforward to show that the points on the local boundary satisfy
	\[1 + (u^2 - 1)^2 - u(2 - |E_3|) + (2 - |E_3| + u - u^3)\sqrt{u^2 - 2} = 0.\]

	\subsection{Tests for network nonlocality close to the $W$ and $\overline{W}$ distributions}
	\label{sec: appendix inequalities W}
	Now we address the regions close to the $W$ and $\overline{W}$ distributions.
	In fact, we just need to study one of them, since the other is obtained from it by performing the transformation $E_1,E_2,E_3\rightarrow -E_1,E_2,-E_3$.
	
	Let us, without loss of generality, focus on the boundary close to the $W$ distribution.
	This boundary is a continuous (but not everywhere smooth) piecewise composition of five surfaces.
	Nonlocal distributions in this region violate {\it all} the inequalities \eqref{ineq W1}-\eqref{ineq W5}.
	In order to obtain them, we have performed the same parameter elimination process described in the previous section to the five families of models that describe the boundary of the set close to the $W$ distribution (i.e., those given by Eqs.~\eqref{W1}-\eqref{W5}).
	For some of them, this can be done in a straightforward manner by using two correlator equations to solve for the degrees of freedom, and then substituting these expressions in the remaining correlator.
	For the remaining ones, the equations result in high-degree polynomials with no explicit solutions involving elementary operations.
	In such cases, we used software for symbolic computation to perform the elimination with the aid of Gr\"obner bases \cite{groebner_basis}.

	The simplest inequality found is that associated with model (\ref{W1}), given by
	\begin{equation}
		-E_1 E_2 + E_3 + (1 - E_1)\sqrt{(1+E_2)^2 - 4E_1^2} \geq 0.\label{ineq W1}
	\end{equation}
	In contrast with Eq.~\eqref{ineq GHZ repeat}, in this case the expression inside the square root in inequality is always nonnegative due to the positivity constraints.
	
	The next inequality is the one associated with model (\ref{W2}).
	It is a polynomial inequality of degree 5, but notice that it is only relevant if an auxiliary condition on $E_1$ is satisfied:
	\begin{align}
		&8 E_1^{5} - 15 E_1^{4} - 16 E_1^{3} E_2 + 22 E_1^{3} + 16 E_1^{2} E_2^{2} - 2 E_1^{2} E_2 - 20 E_1^{2} - 6 E_1 E_2^{2} + 12 E_1 E_2 + 10 E_1 - E_2^{2} - 6 E_2 \nonumber\\
		&\quad+ E_3^{2} \left(E_1^{2} - 2 E_1 + 2\right) + E_3 \left(6 E_1^{3} - 8 E_1^{2} E_2 - 12 E_1^{2} + 10 E_1 E_2 + 10 E_1 - 8 E_2\right) - 1 \geq 0,\label{ineq W2}\\
		&\text{\textbf{provided} } E_1 \geq 1/3.\nonumber
	\end{align}
	If the auxiliary condition $E_1 \geq 1/3$ is false, inequality (\ref{ineq W2}) need not be evaluated and should be considered violated.
	These auxiliary conditions are occasionally necessary because each of the surfaces defining the $W$ boundary has a different range of validity when projected onto the $E_1$-$E_2$ plane.
	
	Related to model (\ref{W3}), we obtain the following inequality of degree 5, with two auxiliary conditions on $E_1$ and $E_2$:
	\begin{align}
		&2 (1 + E_1) E_3^3 + 27 (1 - 3 E_1 + 4 E_1^2 + E_2 - 2 E_1 E_2) (1 - E_1 + 2 E_1^3 + 2 E_2 - 
		E_1 E_2 + E_2^2) \nonumber\\
		&\quad+ 54 E_1 E_3 (2 - 5 E_1 + 5 E_1^2 + 3 E_2 - 4 E_1 E_2 + E_2^2)  
		- 9 E_3^2 (1 - 2 E_1 - 6 E_1^2 + E_2 + 2 E_1 E_2) \geq 0, \label{ineq W3}\\
		&\begin{aligned}
			&\text{\textbf{provided} }&&- 2048 E_1^{12} + 8 E_1^{11} (479 + 128 E_2) - 36 E_1^{10} (153 + 61 E_2) 
			-6 E_1^9 (2033 + 466 E_2 + 177 E_2^2) \nonumber\\
			&&&\quad+ 3 E_1^8 (6687 + 7809 E_2 + 181 E_2^2 + 243 E_2^3) - 4 E_1^7 (2021 + 7383 E_2 + 3159 E_2^2 - 351 E_2^3) \nonumber\\ 
			&&&\quad+ 8 E_1^6 (177 + 3460 E_2 + 264 E_2^2 - 171 E_2^3) - 4 E_1^5 (2303 + 581 E_2 + 10892 E_2^2 - 39 E_2^3 - 621 E_2^4) \nonumber\\
			&&&\quad+ 2 E_1^4 (8625 + 12715 E_2 - 9853 E_2^2 + 9081 E_2^3 + 1728 E_2^4) \nonumber\\
			&&&\quad- 4 E_1^3 (2335 + 11013 E_2 + 8469 E_2^2 - 4405 E_2^3 - 666 E_2^4) \nonumber\\
			&&&\quad+ 4 E_1^2 (125 + 3175 E_2 + 7926 E_2^2 + 5032 E_2^3 - 212 E_2^4 - 108 E_2^5) \nonumber\\
			&&&\quad+ E_1 (506 + 624 E_2 - 2682 E_2^2 - 5100 E_2^3 - 2268 E_2^4) -71 - 369 E_2 - 477 E_2^2 - 243 E_2^3 \geq 0,\nonumber\\ 
			&\text{ \textbf{and} } && 4E_1^2(3 - 2E_1 + E_2 + E_1^2) - (1 - 2E_1)(1 + E_2)^2\geq 0,\nonumber\\
			&\text{ \textbf{and} } && E_3 \geq 6 - 3\sqrt{5}.\nonumber
		\end{aligned}
	\end{align}
	If any of the three auxiliary conditions is false, then inequality (\ref{ineq W3}) should be considered violated.
	
	Next, model (\ref{W4}) leads to a polynomial inequality of degree 8:
	
	\begin{align}
		& (1 - E_3)^4(2 + E_1 + 9 E_1^2 - 2 E_1^3 - 3 E_2 - 
		7 E_1 E_2) + 
		2 (1 - E_3)^3 (E_1 - E_2) (3 E_1 - 17 E_1^2 + 4 E_1^3 + E_2 + 9 E_1 E_2) \nonumber\\
		&\quad- 2 (1 - 
		E_3)^2 \Big(30 E_1^2 - 65 E_1^3 + 43 E_1^4 - 18 E_1^5 - 8 E_2 + 4 E_1 E_2 + E_1^2 E_2 + 21 E_1^3 E_2 + 12 E_1^4 E_2 - 2 E_2^2 + 17 E_1 E_2^2 \nonumber\\
		&\quad- 43 E_1^2 E_2^2 - 2 E_1^3 E_2^2 - E_2^3 + 11 E_1 E_2^3\Big) - (1 - E_3)\Big(16 E_1^2 - 128 E_1^3 + 261 E_1^4 - 227 E_1^5 + 88 E_1^6 - 
		16 E_2 + 80 E_1 E_2 \nonumber\\
		&\quad - 300 E_1^3 E_2 + 428 E_1^4 E_2 - 232 E_1^5 E_2 - 
		48 E_2^2 
		+ 128 E_1 E_2^2 - 58 E_1^2 E_2^2 - 130 E_1^3 E_2^2 + 168 E_1^4 E_2^2 - 32 E_2^3 + 68 E_1 E_2^3 \nonumber\\
		&\quad - 36 E_1^2 E_2^3 - 40 E_1^3 E_2^3 - 3 E_2^4 + 13 E_1 E_2^4\Big) -(1 - E_3)^5(1 + E_1)
		+ 48 E_1^3 - 6 E_1^4 - 
		515 E_1^5 + 1101 E_1^6 - 882 E_1^7 \nonumber\\
		&\quad + 256 E_1^8 - 48 E_1 E_2 + 48 E_1^2 E_2	+ 280 E_1^3 E_2 - 467 E_1^4 E_2 - 71 E_1^5 E_2 + 504 E_1^6 E_2 - 256 E_1^7 E_2 + 48 E_2^2  \nonumber\\
		&\quad - 224 E_1 E_2^2 + 220 E_1^2 E_2^2 + 298 E_1^3 E_2^2 - 598 E_1^4 E_2^2 + 212 E_1^5 E_2^2 + 64 E_1^6 E_2^2 + 64 E_2^3 - 
		200 E_1 E_2^3 \nonumber\\
		&\quad + 82 E_1^2 E_2^3 + 218 E_1^3 E_2^3 
		- 184 E_1^4 E_2^3 + 
		26 E_2^4 - 39 E_1 E_2^4 - 7 E_1^2 E_2^4 + 30 E_1^3 E_2^4 + E_2^5 - 3 E_1 E_2^5 \geq 0, \label{ineq W4}\\
		&\text{\textbf{provided} } E_1-E_2-E_3 \leq 29/27.\nonumber
	\end{align}
	The auxiliary condition for inequality (\ref{ineq W4}) has an interesting geometric interpretation. It is the halfspace defined by the plane containing the distribution generated by model (\ref{W example}) and parallel to the affine subspace of mixtures of $D_+$, $D_-$, and $\overline{W}$. Because of that, it can be used as a primary check that eliminates a sizeable portion of the $W$ region, thereby eliminating the need of performing the complete $W$ test in many cases.
	
	Finally, we give the inequality related to model (\ref{W5}) expressed as
	\begin{equation}
		E_3 \geq f(E_1, E_2),\label{ineq W5}
	\end{equation}
	where $f$ is a complicated function, described in appendix \ref{sec: appendix optimization}.
	
	It is important to recall that the test for network nonlocality entails checking the violation of all Eqs.~\eqref{ineq W1}-\eqref{ineq W5}.
	If a distribution does not violate all of them, then it is not in the nonlocal region close to the $W$ distribution, and might even be local, if it does not pass the remaining tests.
	
	\section{Characterization of model (\ref{W5}) and inequality (\ref{ineq W5})}
	\label{sec: appendix optimization}
	
	With the exception of model (\ref{W2}), all the other models that characterize distributions in the boundary close to the $W$ distribution are particular cases of the following general 333 model, up to hidden variable exchanges and relabellings:
	\begin{align}
		&q =
		\begin{bmatrix}
			x & y & 1-x-y
		\end{bmatrix}, \, r = 
		\begin{bmatrix}
			z & t & 1-z-t
		\end{bmatrix},\,
		s = 
		\begin{bmatrix}
			u & v & 1-u-v
		\end{bmatrix},\, A = 
		\begin{bmatrix}
			0 & 1 & 0\\
			1 & 1 & 0\\
			1 & 1 & 1
		\end{bmatrix},\, B = 
		\begin{bmatrix}
			0 & 1 & 1\\
			0 & 0 & 1\\
			1 & 1 & 1
		\end{bmatrix},\, C = 
		\begin{bmatrix}
			w & 1 & 1\\
			1 & k & 0\\
			1 & 0 & 0
		\end{bmatrix},\label{W general}
	\end{align}
	subject to the positivity constraints
	\begin{equation}
		x\geq 0,\,\, y\geq 0,\,\, x+y \leq 1,\,\, z\geq0,\,\, t\geq0,\,\, z+t\leq 1,\,\, u\geq0,\,\, v\geq0,\,\, u+v \leq 1,\,\, 0\leq w\leq 1,\,\, 0\leq k\leq 1.\label{positivity_constraints}
	\end{equation}
	This model has four degrees of freedom (eight parameters minus four symmetry constraints), and it spans a volume in $(E_1,E_2,E_3)$ space.
	In order to find the boundary of this volume closest to the $W$ distribution, it suffices to solve the constrained optimization problem of finding the combination of parameters that minimizes $E_3$ for fixed values of $E_1$ and $E_2$, while satisfying the positivity and symmetry constraints.
	By numerically solving this problem, we found which positivity constraints  saturate depending on the values of $E_1$ and $E_2$.
	This way, the general model (\ref{W general}) splits into four different subclasses.
	For each subclass, the problem of minimizing $E_3$ is much more manageable and can be tackled analytically, yielding the models presented in appendix \ref{sec: appendix models}:
	
	\begin{center}
		\begin{tabular}{c|c}
			Saturated positivity constraints~\eqref{positivity_constraints} & Model\\
			\hline
			$x+y=1$ and $z+t=1$ & (\ref{W1})\\
			no constraints saturated & (\ref{W3})\\
			$z+t=1$ and $k=1$ & (\ref{W4})\\
			$k=1$ & (\ref{W5})
		\end{tabular}
	\end{center}
	
	Models (\ref{W1}), (\ref{W3}) and (\ref{W4}) were already completely described, but for model (\ref{W5}) the solution is too complicated to put in print (or even to calculate explicitly, for that matter), so the goal of this appendix is to offer an implicit description of it.
	In the process, we will also specify the function $f$ that appears in inequality (\ref{ineq W5}).
	
	By enforcing the four symmetry constraints and the two equality constraints for $E_1$ and $E_2$, we can express the parameters $y$, $t$, $u$, $v$, $w$, $k$ as functions of $x$ and $z$.
	As a result of that, $E_3$ is also expressed as a function of these parameters, and we just have to minimize $E_3(x,z)$ subject to $k(x,z) = 1$.
	The functions $E_3$ and $k$ are given by:
	\begin{align*}
		E_3(x,z) &= \frac{P(x,z)}{2xz(1+E_1-x-z)},\\
		k(x,z) &= \frac{Q(x,z) R(x,z) S(x,z)}{4xz(1+E_1-x-z)(1-2E_1+E_2)T(x,z)U(x,z)},
	\end{align*}
	where the factors $P$, $Q$, $R$, $S$, $T$, and $U$ are:
	\begin{align*}
		P &= (1+E_1-2z)^2(1-2E_1+E_2) - 2x\Big[2(1+E_1)(1-2E_1+E_2) + z(x+z)(3-11E_1 - 2E_1^2 + 3E_2) \nonumber\\
		& \quad- z(5-10E_1+5E_2-9E_1^2+3E_1E_2) - 2x(1 - 2E_1 + E_2 - 4z^2E_1)\Big],\\
		Q &= - 4 x^{2} z E_1 + 4 x^{2} z + 4 x z^{2} E_1 + 4 x z^{2} - 4 x z E_1 - 4 x z + 4 x E_1 - 2 x E_2 - 2 x - 2 E_1^{2} + E_1 E_2 - E_1 + E_2 + 1,\\
		R &= 4 x^{2} z E_1 + 4 x^{2} z - 4 x z^{2} E_1 + 4 x z^{2} - 4 x z E_1 - 4 x z + 4 z E_1 - 2 z E_2 - 2 z - 2 E_1^{2} + E_1 E_2 - E_1 + E_2 + 1,\\
		S &= (1+E_1 - 2z)(1+E_1)(1 - 2E_1 + E_2) - 2x\Big[(1+E_1)(1 - 2E_1 + E_2) - 8 x z^{2} E_1 + 2 x z E_1^{2} + 4 x z E_1 + 2 x z E_2 \nonumber\\
		&\quad + 2 z^{2} E_1^{2} + 4 z^{2} E_1 + 2 z^{2} E_2 - 2 z E_1^{2} - 2 z E_1 E_2 + 2 z E_1 - 4 z E_2 - 2 z\Big],\\
		T &= (1+E_1)(1-2E_1+E_2) + 4xz(2 -z - x + zE_1 - xE_1) + 4 x E_1 - 2 x E_2 - 2 x - 4 z^{2} E_1 + 4 z^{2} + 4 z E_1^{2} - 4 z,\\
		U &= (1+E_1 -2z)(1 - 2E_1 + E_2) -4 x + 4 x^2 + 8 x z - 4 x^2 z - 4 x z^2 - 4 x^2 E_1 + 4 x^2 z E_1 - 
		4 x z^2 E_1 + 4 x E_1^2.
	\end{align*}
	
	Next, the optimal values of $x$ and $z$ can be found with the method of Lagrange multipliers, resulting in a system of two equations:
	\begin{equation}
		\label{system}
		\begin{aligned}
			\frac{\partial E_3}{\partial x}\frac{\partial k}{\partial z} &= \frac{\partial E_3}{\partial z}\frac{\partial k}{\partial x},\\
			k(x,z) &= 1,
		\end{aligned}
	\end{equation}
	which can be expressed as two polynomial equations of degrees 18 and 10 in the variables $x$ and $z$.
	These polynomials have too many terms to find a solution via Gr\"obner basis in a reasonable amount of time, so we solved the system numerically in order to generate Fig. \ref{fig:3d-zoom}.
	In general, there are multiple solutions for (\ref{system}), so one has to check which ones satisfy the positivity constraints and then select, from those, the pair $(x^*, z^*)$ that yields the smallest value of $E_3$. This completes the description of model (\ref{W5}).
	
	The function $f$ mentioned in inequality (\ref{ineq W5}) is exactly the value of $E_3$ resulting from this minimization procedure:
	\begin{equation}
		f(E_1, E_2) \coloneq E_3(x^*, z^*),
	\end{equation}
	with the proviso that if system (\ref{system}) has no feasible solution, then inequality (\ref{ineq W5}) should be considered violated.

	\section{Proof that inequality (\ref{ineq GHZ}) recovers the NSI boundary}
	\label{sec: appendix nsi saturation}

	In this appendix we show that inequality (\ref{ineq GHZ}) saturates the NSI boundary given by inequality (15) of Ref. \cite{NSI} when $E_3\leq-2+\sqrt{2}$ and $E_1+E_2-E_3=1$. This region is depicted in Fig.~\hyperref[fig: 2d subspaces]{2(b)}. 
	
	Substituting $E_3 = E_1 + E_2 - 1$ into inequality (\ref{ineq GHZ}) and factoring the resulting polynomials with the positivity constraint $-1\leq E_1 \leq 1$ in mind, we get:
	\begin{align*}
		16(1+E_1)^4 \Big[\sqrt{2 - 2E_1}(3 - 4E_1 + E_2) - 4 + 8E_1 - 2E_2 - 2E_1^2\Big] \geq 0.
	\end{align*}
	
	The factor $16(1+E_1)^4$ is always positive (except for $E_1 = -1$, in which case the remaining factor also evaluates to zero).	Thus, we can just focus on the positivity of the second factor, which reads
	\begin{equation}
		\sqrt{2 - 2E_1}(3 - 4E_1 + E_2) \geq 4 - 8E_1  + 2E_2 + 2E_1^2.\label{proving_saturation}
	\end{equation}
	 
	Next, we proceed by proving that 
	 the right hand side of this inequality is non-negative for valid behaviors, in which case we can square both sides of inequality (\ref
	 {proving_saturation}). All valid probability distributions satisfy $p(a=-1,b=-1) = (1 - 2E_1 + E_2)/4 \geq 0$, yielding the positivity constraint $E_2 \geq 2E_1 - 1$. Thus, the right hand side of inequality (\ref{proving_saturation}) can be bounded as follows:
	 \[4 - 8E_1 + 2E_2 + 2E_1^2 \geq 4 - 8E_1 + 2(2E_1 - 1) + 2E_1^2 = 2 - 4E_1 + 2E_1^2 = 2(1 - E_1)^2 \geq 0.\]
	 Therefore, we can square both sides of inequality (\ref{proving_saturation}), which can then be factored as
	 \[2(1+E_1)(1 - 2E_1 + 2E_1^2 - 2E_1^3 - 2E_2 +4E_1E_2 - E_2^2) \geq 0.\]
	 Taking into account the positivity constraints $-1 \leq E_1 \leq 1$ and $-1/3 \leq E_2 \leq 1$, we can divide both sides by $2(1+E_1)$:
	 \begin{align*}
		&1 - 2E_1 + 2E_1^2 - 2E_1^3 - 2E_2 + 4E_1E_2 - E_2^2 \geq 0,\\
		& 2 - 6E_1 + 6E_1^2 - 2E_1^3 \geq 1 - 4E_1 + 4E_1^2 + 2E_2 - 4E_1E_2 + E_2^2,\\
		& 2(1 - E_1)^3 \geq (1 - 2E_1 + E_2)^2,
	 \end{align*}
	 which is the NSI bound characterized by inequality (15) of Ref. \cite{NSI} when $E_1 \leq 0$.	This means that inequality (\ref{ineq GHZ}) does in fact describe the true network-local boundary for the region depicted in Fig.~\hyperref[fig: 2d subspaces]{2(b)} when $E_3 \leq -2 + \sqrt{2}$.
\end{document}